\documentstyle[manuscript,aps,prb]{revtex}
\begin{document}
\draft
\input psfig
\draft
\title{Charge ordering, phase separation and charge pairing in layered 3D systems}
\author{Yu.G.Pashkevich\footnote{Corresponding author, e-mail:pashkevi@kinetic.ac.donetsk.ua} and A.E.Filippov}
\address{Donetsk Phystech National Academy of Sciences of Ukraine, 340114 Donetsk,
Ukraine}
\date{\today }
\maketitle

\begin{abstract}
The processes of Coulomb gas ordering in 3D layered system are studied by
means of Brownian dynamics approach. It is found that at different densities
of the carriers the 3D lattice of charges as well as new specific structures
are possible in the system. In particular, at some small density the
particles inside the layers can associate into droplets that collectively
repulse between neighbouring layers. These droplets possess local stripe
structure which orders spontaneously along arbitrary chosen direction. At
higher densities a specific ordering of the charges into the tetragonal-like
or hexagonal-like structures is observed visually and described numerically.
Specific ''pairing'' of the charges from different layers plays an essential
role in formation of all above structures. It is essentially many-body
effect in 3D system which can be a reason for unusual properties of layered
crystals.
\end{abstract}


\section{Introduction}

The theory of doped Mott insulator systems have shown variety of electronic
ground states from electronic liquid crystal phases \cite{kfe} to the usual
metallic state that depend on doping level. The complex character of phase
diagram is caused by the long-range Coulomb interaction between electrons
placed in an antiferromagnetic background. Besides, the well defined layered
structure is a common feature of these systems at all variety of other
properties. Both these features are essential difficulties that obstacle to
construct correct theory here up to the moment.

Therefore, the majority of the works in the field are the studies of the
{\it two}-dimensional systems with quite complicated interactions. The
separation of the charge and spin variables is an additional simplification
here \cite{Bishop,Berlinsky}. Sometimes such models are not quite realistic
and can not be applied to describe the systems without magnetic ions. For
example, some systems without magnetic ions, like Ba$_{1-x}$K$_x$BiO$_3$,
have shown a remnant of charge ordering above T$_c$ -temperature of
superconductive phase transition for the optimally doping value of x=0.4.

In present article we show that the density stratification (or phase
separation) accompanied by a collection of classic phases (like Wigner
crystal or stripe-ordering) appears already in a very simple model of three
dimensional layered crystal with long-range Coulomb interaction among
spinless charges.

One deals here with a case that is extremely difficult for the theory. From
one side the well pronounced anisotropy and layered structure do not give a
chance to limit ourselves by simple mean-field approximation which is
suitable for analytic theory. From another one, the essentially many-body
and 3D character of the system makes it difficult to simulate its behavior
directly.

However, the development of the computer technology continuously lowers the
extent of the simplification that is necessary for numerical simulation. As
a rule, the simulation in two or three dimensions is performed separately.
Keeping in mind an application to above layered structures, it is
interesting to simulate just intermediate case. Let us suppose that the
equally charged spinless particles move and interact in 3D space with a
positive background that includes a periodic potential along one of the
dimensions which we choose as $z$-direction.

First of all, let us discuss briefly qualitative effect of ''layer
structure''. At some conditions the particles should be found as localized
completely in a vicinity of the equidistant layers. It generates a strong
impact from the 2D behavior into collective behavior. The particles tend to
produce a hexagonal (Wigner) crystal in each layer.

The particles from neighboring layers interact with collective potential of
this crystal and tend to dispose in its minima. In an ideal (translational
invariant) case the particles have to be found under centers of the
neighboring layer exactly. The particles from a next layer should be found
just under positions of the particles from the first one. However, these
particles repulse too. So, the energy minimum, should be a nontrivial
compromise between long-range interaction of the particles and the energy of
bonding inside the layers.

Numerical simulation shows that the symmetric position of the particles over
the centers of the cells of the Wigner crystal is unstable. They are shifted
(in a projection onto layer plane) in a direction of one of the neighboring
particles. In particular, it breaks the geometric frustration in the
collective position of a hexagonal 2D lattice from one layer under another.

It looks like as follows. Moving in a collective field of the neighboring
layers the carriers attract to the minima of the potential. These minima
play a role of effective ''positively charged particles''. The projections
of real particles from the neighboring layers (attracted to these minima) on
the $xy$-plane can be treated as ''images'' of these effective particles.
This is essentially 3D picture, but in some sense it is quite close to a
stripe formation in cuprates and nikelates \cite{kfe,zaa98}. It is direct
analogy to previously studied process of screening in a system containing
two kinds of the particles that bind into ''pairs''. These pairs are
dipolarly charged and form, in their turn, the chains of the dipoles \cite
{Filippov}.

Let us note once more that the ''pairs'' mentioned above are formed by
particles from different layers. So, this pairing is essentially 3D
phenomenon. Moreover, we found that the variation of the density leads to
other nontrivial structures also. In particular, at low density the system
produces an unusual ''droplet phase''. This phase consists of the charged
droplets with an internal structure. One can believe that the specific
dipolar chain structure or structuralised droplet phase can attitude to
actively studied now mechanism of superconductivity of novel superconductors.

\section{Numerical simulation}

To restore the structures appearing in the system of moving charges placed
into 3D compensating background with periodically modulated potential along
one of the dimensions we apply the Brownian dynamics (BD) technique. This
technique is widely applied last years \cite
{Persson,Elmer,Urbakh1,Urbakh2,Braun1,Braun2}\ (in particular, by the
authors of present article \cite{BraunHFZ,VakFBad}) to simulate a behavior
of different systems. It is useful in the cases when the discrete
description in terms of the moving particles is more natural than the
continuous kinetic approach.

The technique is based on the solution of the system of the dynamic
equations for the particles with discrete set of the coordinates ${\bf R}%
_j=\{x_j,y_j,z_j\}$ where $1\leq i\leq N$.and the vectors ${\bf R}_{jk}={\bf %
R}_j-{\bf R}_k=\{x_j-x_j,y_j-y_j,z_j-z_j\}$ connecting different particles.
The choice of appropriate boundary conditions depends on the specific
problem.

Mutual relation between this and kinetic approaches is obtained due to
following two features of the description. First of all, the equations in
such an approach contain some random noise sources \cite
{Braun1,Braun2,BraunHFZ,VakFBad}. Together with a relaxation terms in the
equations this reproduces an effect of final (nonzero) temperature (in
according to known fluctuation-dissipative theorem). Besides, if it is
necessary, the continual densities of all values of the problem are restored
''$a\;posteriory$'' by means of a summation over realizations and averaging
during sufficiently long time. This time is also determined from the
numerical simulation for a specific problem \cite{BraunHFZ,VakFBad}.

In particular the density of particle distribution in real space $\varrho _j(%
{\bf R})=<<\sum_i\delta ({\bf R}-{\bf R}_j)>>\mid _{t_0}$is restored by a
summation $\sum_i\;$over particle coordinates and averaging $<<...>>\mid
_{t_0}$\ over characteristic time $t_0$. This density is taken to calculate
the correlation functions of the problem (for example: two-point correlation
function $G({\bf R},{\bf R}^{\prime })=\langle \varrho ({\bf R})\varrho (%
{\bf R}^{\prime })\rangle $ , where $\langle ...\rangle $ denotes an
averaging over ensemble of particles). The set of the correlation functions
gives complete information about the thermodynamic properties of the system.

More detail description of the approach, as well as some examples of its
utilization, can be found in previous publications by different authors \cite
{Persson,Elmer,Urbakh1,Urbakh2,Braun1,Braun2}. In present work , we limit
ourselves by relatively short positing of the problem and discussion of
numerical results.

The set of BD equations can be written in a form

\[
\ddot{{\bf R}_j}+\gamma \dot{{\bf R}_j}+\frac \partial {\partial {\bf R}_j}V(%
{\bf R}_j)+\frac \partial {\partial {\bf R}_j}\sum_kU({\bf R}_{jk})=\delta
\!F({\bf R}_j;t)
\]
Here $\gamma $\ is relaxation constant. To model a thermal bath we apply the
Gaussian random force $\delta \!F(t)$,
\begin{eqnarray*}
\langle \delta \!F({\bf R};t)\rangle &=&0; \\
\langle \delta \!F({\bf R};t),\delta \!F({\bf R}^{\prime };t^{\prime
})\rangle &=&2\gamma T\delta ({\bf R}-{\bf R}^{\prime })\delta (t-t^{\prime
})
\end{eqnarray*}

Periodic one-particle potential is chosen in following form

\[
V({\bf R}_j)=V_0\cos (2\pi k{\bf R}_j/L_z)
\]

The system is supposed to be continued periodically. Here $L_z/k$ is a
period of the potential along the $z$-axis and $k$ is an integer number.
Along two other axes the system is quasi-infinite. The boundary conditions
along all the axes are as follows. Every particle interacts with the
particles inside the calculation volume $\Omega =L_x\cdot L_y\cdot L_z$ as
well as with all their ''images'' obtained as a result of the translation of
particle coordinates across (beyond) nearest boundary \cite
{Braun1,Braun2,BraunHFZ,VakFBad} . So, the many-body potential $U({\bf R}%
_{jk})\;$in the equations contains the vectors $x_{jk}=x_{jk}\pm
L_x,y_{jk}=y_{jk}\pm L_y,z_{jk}=z_{jk}\pm L_z$ with all possible
permutations. The sign ''+'' or ''-'' is determined by a direction to the
nearest boundary plane along given axis. Besides, if moving particle leaves
the volume $\Omega =L_x\cdot L_y\cdot L_z$ along one of the axes it is
returned to the volume. At absence of external field it can by done or by
means of mirror reflection, or by cyclic shift. In the second case
corresponding projection of the velocity conserves and the coordinate
projection is shifted as follows: $x_j\rightarrow x_j\pm L_x,y_j\rightarrow
y_j\pm L_y,$ $z_j\rightarrow z_j\pm L_z$ \cite{BraunHFZ}.

This approach is quite suitable for the simulation of translation invariant
system of charges in solid state. Used here many body potential $U({\bf R}%
_{jk})=U_{coulomb}({\bf R}_{jk})\cdot U_{screen}({\bf R}_{jk})\;$contains
from long-range coulomb interaction $U_{coulomb}({\bf R}_{jk})\;\symbol{126}%
\;1/\mid {\bf R}_{jk}\mid $ and cut-off screening impact\ $U_{screen}({\bf R}%
_{jk})\symbol{126}\exp (-\mid {\bf R}_{jk}\mid /r_0)$ from crystal lattice
and other subsystems of the problem.

\begin{figure}\centerline{\psfig{figure=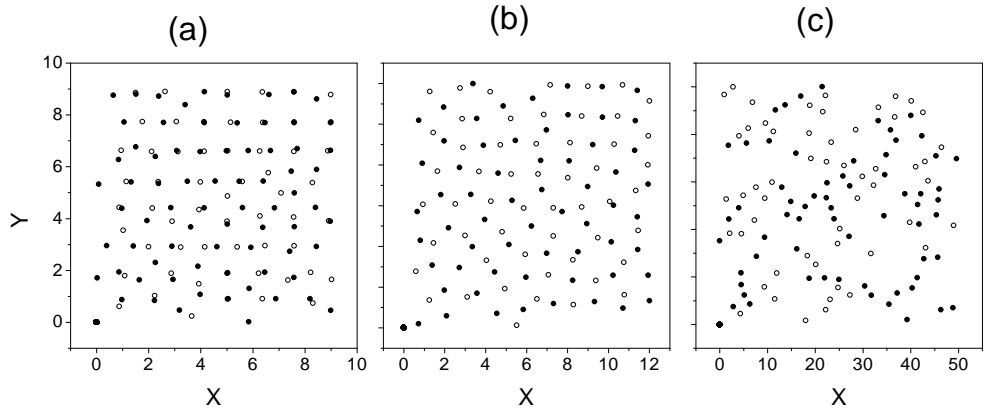,width=13cm}}
\caption{ \label{Snap1}Instant configurations of the particles at
final stage of evolution are shown for the three characteristic
densities: (a) at $L_x=L_y=9$ ; (b) at $L_x=L_y=12$ ; (c) at
$L_x=L_y=50$ ($N=256)$. The projections of the particles from
different layers are shown by black and white circles
respectively.}
\end{figure}

\begin{figure}\centerline{\psfig{figure=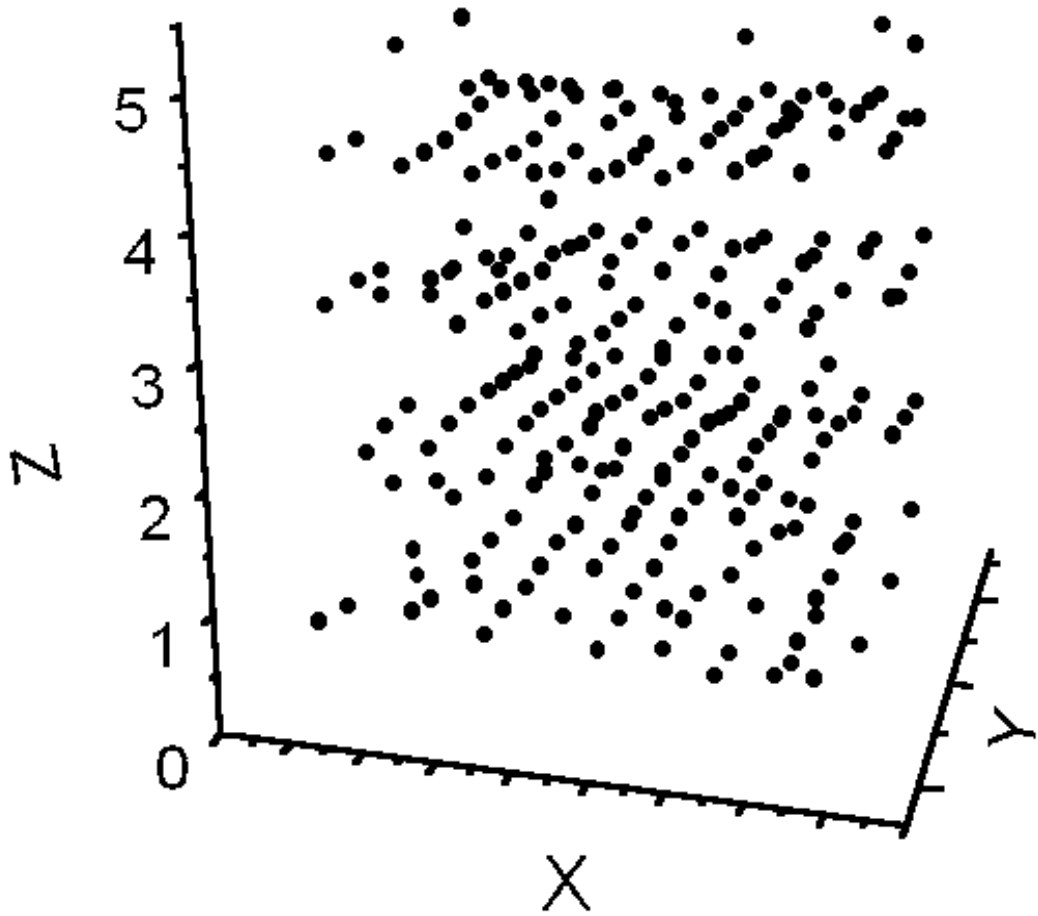,width=5.4in}}
\caption{ \label{Tetra}
Typical tetragonal 3D structure appearing at high density ($%
\;L_x=L_y=9 $ ). Long-range waves of z-displacements of particles caused by
a strong inter-layer interaction are seen directly.}
\end{figure}

\begin{figure}\centerline{\psfig{figure=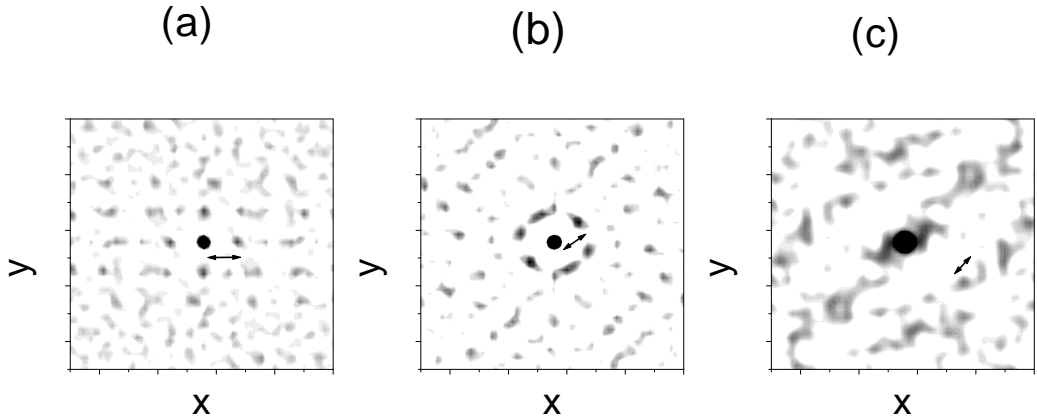,width=5.4in}}
\caption{ \label{corr} Two-point correlation function $G({\bf
R},{\bf R}^{\prime })=\langle \varrho ({\bf R})\varrho ({\bf
R}^{\prime })\rangle $ is presented for following three typical
cases: (a) well pronounced tetragonal lattice (at $L_x=L_y=7$);
(b) for hexagonal ordering of the particles inside the layers
($L_x=L_y=10.5$); (c) for droplet phase with an internal stripe
structure appearing at phase separation (at small density
$L_x=L_y=50$). The length of arrows indicate the value of the
averaged minimal distance between particles for given phase.}
\end{figure}

\begin{figure}\centerline{\psfig{figure=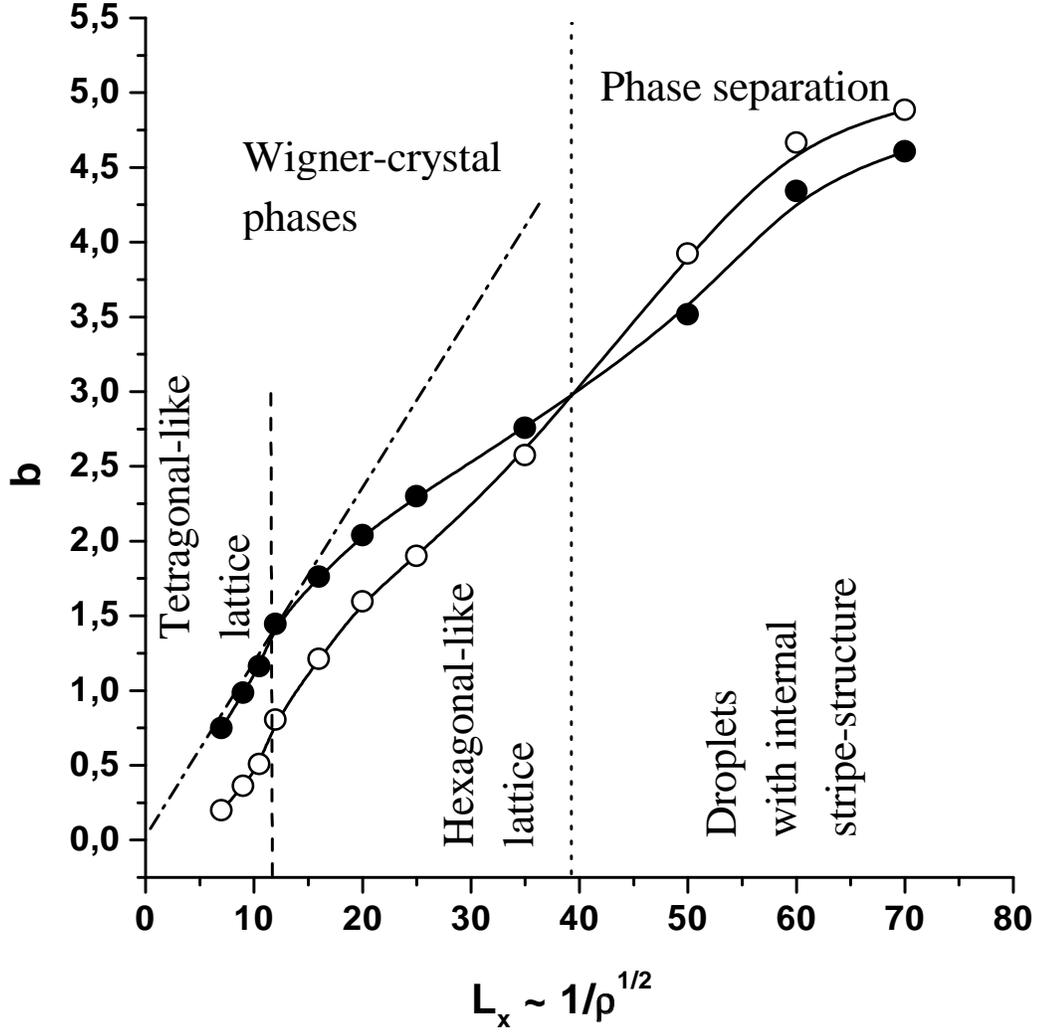,width=5.4in}}
\caption{ \label{Phase} A dependence of equilibrium averaged
minimal distance between the particles $b=<\min \{\mid {\bf
R}_{jk}\mid \}>$\ from density (shown as a function from box size
$L_x=L_y\;$ at fixed number of particles $N=256$). Two kinds of
the circles denotes the values of $b$ corresponding to the
particles from the same layers (black circles) or different layers
(white circles). Sloping dash-doted line gives a slope of the
$b(L_x)$ function for an ideal tetragonal lattice. Vertical dashed
line (at $L_x\approx 11$) denotes a transition from tetragonal to
hexagonal ordering. Dotted vertical line corresponds to a
transition from the ordering of separate charges to the
concentration stratification.}
\end{figure}

Numerical results are summarized in the Figs.\ \ref{Snap0}-\ref{Phase}. For
definiteness all instant configurations of the particles here were obtained
in the frames of uniform kinetic scenario. The random distributions of the
charges over system have been taken as the initial conditions. From physical
point of view appearance of such carriers in the system can be treated as a
result of a doping. In its turn, it corresponds to standard initial
conditions for a search of an energy minimum by relaxation technique \cite
{L-Khal} and annealing technique \cite{Bishop,Loskutov}.

Besides, equal number of the particles $N$=256 has been used for all the
cases. A number of the variables of 3D problem (the coordinates and
velocities) corresponding to this number of particles is equal to $%
2dN=1536\approx 1.5\cdot 10^3$. In its turn, it corresponds to approximately
$2.5\cdot 10^6$ of their mutual combinations that should be calculated at
every step. At given number of the particles $N$=256 the charge density is
varied by change of the cross-section $L_x\cdot L_y$\ of the calculation
volume $\Omega =L_x\cdot L_y\cdot L_z\;$(at periodic boundary conditions
described above). To control the results were reproduced at some densities
for other numbers of particles ($N=128$ and $N=512$\ , with respectively
chosen cross-section of the box $L_x\cdot L_y\;$).

\section{Results and discussion}

At initial stage of the evolution the particles spontaneously group to the
vicinities of the planes corresponding to the minima of periodic potential $%
V({\bf R}_j)=V_0\cos (2\pi k{\bf R}_j/L_z)$. These minima below are named as
the ''layers''. For definiteness, we are denote the particles as belonging
to a given layer in the case if the $z$-coordinate is away from the layer
not more than 0.1 of distance between the layers $\mid z-z_k\mid <0.1\cdot
\mid z_{k+1}-z_k\mid $.

\begin{figure}\centerline{\psfig{figure=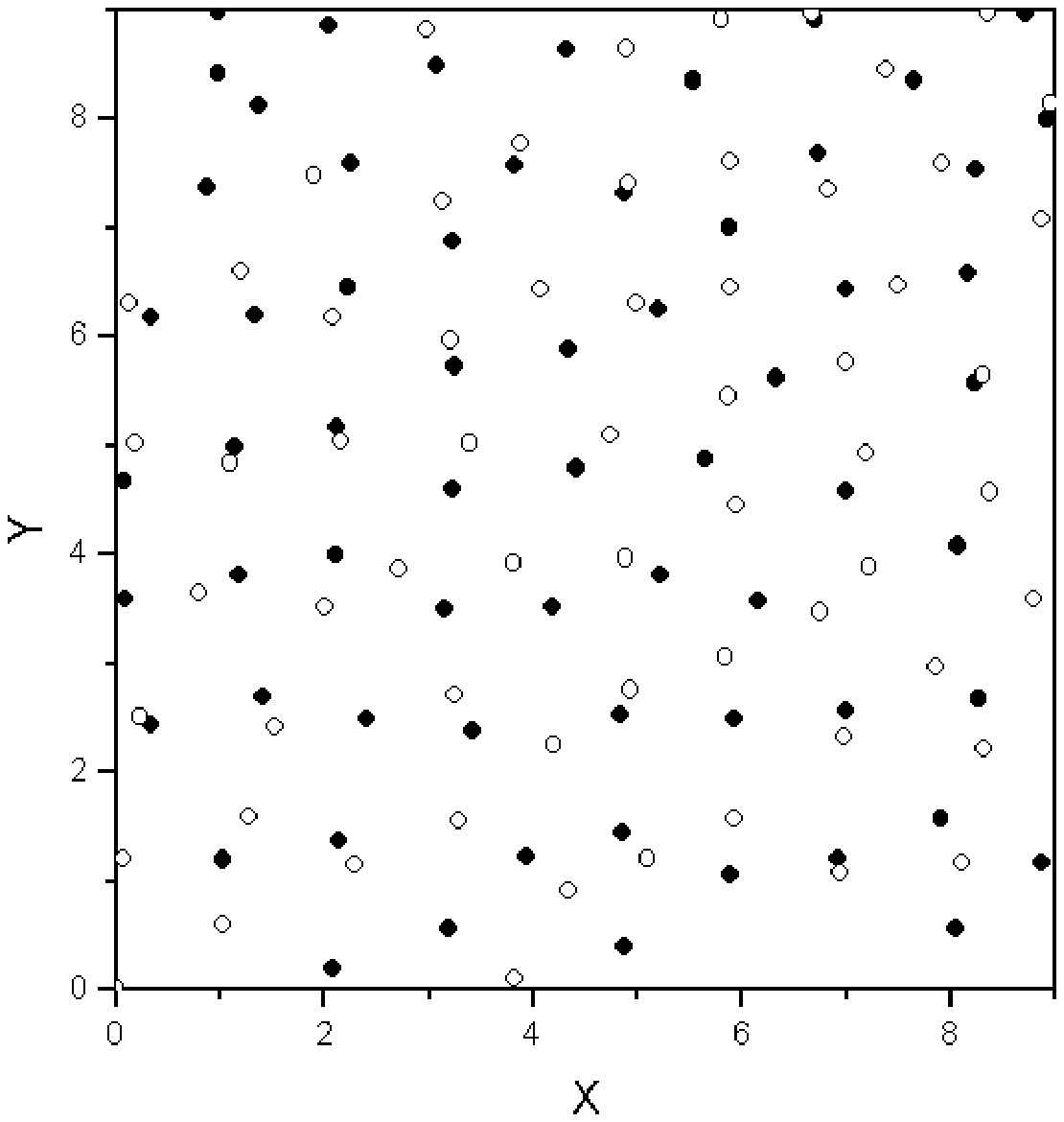,width=13cm}}
\caption{ \label{Snap0}Typical instant intermediate particle
configuration for two neighbouring layers, projected into
$xy$-plane. The particles from different layers are shown by
different circles (black and white respectively). High density
case corresponding to tetragonal equilibrium structure
($L_x=L_y=9$ at $N=256$) is presented.}
\end{figure}

Typical projection on $xy$-plane of an instant configuration of two
arbitrary neighboring layers at early evolution stage is shown in the Fig.%
\ref{Snap0} . The particles from different layers are shown by two kinds of
the circles (black and white respectively). All other particles are not
shown in this figure. Typical for this stage tendency to a ''pairing'' (in
projection onto $xy$-plane) of the particles from neighboring layers is seen
directly.

Below this tendency will be characterized numerically. Let us note that such
a state conserves for some of intermediate densities only. In particular,
the configuration shown in Fig. \ref{Snap0}\ is interesting because it takes
place in system with relatively high density ($L_x=L_y=9$ at $N=256$) at
initial kinetic stage. Another, tetragonal, structure is found to be natural
for this system in a final equilibrium state.

Instant particle configurations at final (but not equilibrium) stages of
structure forming is shown in Fig. \ref{Snap1} for three specific densities.
As before, the only particle coordinates for two consistent layers are shown
by means of different circles in projection onto $xy$-plane. The case (a)
corresponds to already mentioned high density ($L_x=L_y=9$). The
configuration (b) conserves at long time for the density close to some
critical one ($L_x=L_y=12$). And at small density ($L_x=L_y=50$) a kind of
clasterisation into ''droplets'' happens in the system. These droplets
contain of scraps of charge chains and, therefore, possess some kind of fine
structure. The droplets from different layers repulse mutually. The state
obtained looks like analogous distributions obtained at concentration
stratification (or phase separation) of the particles of different kind \cite
{Filippov}. For briefness it could be called as ''phase separation'' too.

One can use a time dependence of minimal distance between particles $b=<\min
\{\mid {\bf R}_{jk}\mid \}>$\ as a numerical characteristics of different
scenarios at different densities $\varrho $. This value corresponds to mean
distance between nearest neighbors. Different mutual relations and sign of
inequality between values of $b$ inside the layers and for the pairs of
neighboring layers reflects a difference between visually observed
structures.

\begin{figure}\centerline{\psfig{figure=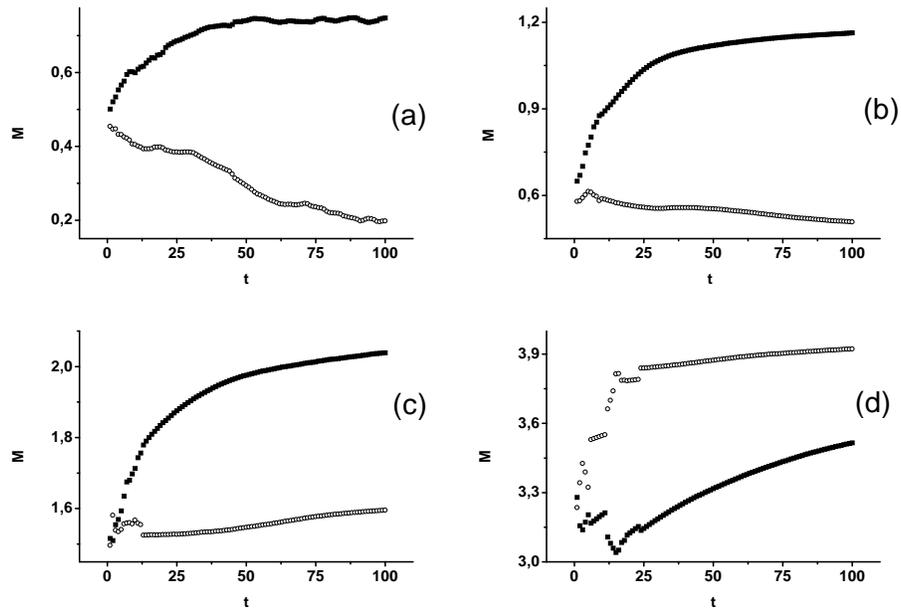,width=13cm}}
\caption{ \label{Time} Four qualitatively different evolution
scenaria of averaged minimal distance between the projections of
particles from the same layer (black
circles) and nearest layers (white circles): (a) at $L_x=L_y=9$ ; (b) $%
L_x=L_y=10.5;$(c) at$\;L_x=L_y=11$ ; (d) at$\;L_x=L_y=70$ ($N=256)$. }
\end{figure}

Four qualitatively different scenarios of evolution of the distance between
nearest particles are shown in Fig.\ref{Time}. It is done for following
cases: (a) at high density ($L_x=L_y=9$); (b) at some density close, but
slightly higher than a critical one ($L_x=L_y=10.5$); (c) (b) at a density
close, but lower than a critical one ($L_x=L_y=11$); and finally (d) at very
low density ($L_x=L_y=70$). The averaged projections of the distances inside
the layers and between them are shown by black and white circles
respectively.

At fixed distance between the layers the average interaction between the
layers is proportional to the density. When the density is high enough this
interaction is quite strong to order particles in 3D structure. The Fig.\ref
{Tetra}\ presents typical tetragonal crystal structure that is obtained at
high density. The particular case of ($L_x=L_y=9$) is shown in isometric
projection. Long-wave $z$-displacements of the particles inside each of
layer are directly visible. It is caused by strong interaction along $z$%
-axis which does not allow to create an ideal 3D tetragonal lattice.

When the density goes to some critical one ($10.5<L_x=L_y<11$) the force
bonding the particles in the layers prevail the $z$-component of the
interaction between layers. The particles are locked strongly in the layers
and ordered hexagonally. Naturally, the mean distance between nearest
neighbors stabilizes with a time. However, the distance between the
projections of particles from different layers does not tend to zero now (as
it takes place for 3D tetragonal lattice). This fact is reflected by
difference in behavior of the lower lines in the Figs.\ref{Time}(a-c).

Finally, when the density is extremely low the repulsion between the
particles from different layers is notable for sufficiently big groups of
the particles. Phase separation occurs now. The particles combine into
clusters (or droplets). The projections of the clusters from different
layers mutually fill in free spaces for each other. A tendency to form
stripes or chains of charge along one of spontaneously chosen direction is
observed in the system.

All these modifications of the structure are reflected by physically
measurable correlation function $G({\bf R},{\bf R}^{\prime })=\langle
\varrho ({\bf R})\varrho ({\bf R}^{\prime })\rangle $ shown in the Fig.\ref
{corr}(a-c). The value $G({\bf R},{\bf R}^{\prime })=\langle \varrho ({\bf R}%
)\varrho ({\bf R}^{\prime })\rangle $ is calculated here for one
representative layer and shown for following three typical cases: (a) well
pronounced tetragonal lattice (at $L_x=L_y=7$); (b) for hexagonal ordering
of the particles inside the layers ($L_x=L_y=10.5$); (c) for droplets phase
appearing at phase separation into charge rich and charge poor domains (at
small density $L_x=L_y=50$). The well defined filament structure caused by
the scraps of charge chains are visible. In all pictures the averaged
minimal distance between particles inside the layer for the given $L_x=L_y$
are marked by the arrows.

A dependence of equilibrium averaged minimal distance between the particles $%
b=<\min \{\mid {\bf R}_{jk}\mid \}>$\ from density (shown as a function from
box size $L_x=L_y\;$at fixed number of particles $N=256$) is summarized in
Fig. \ref{Phase}. As before, two kinds of the circles denotes the values of $%
b$ corresponding to the particles from the same or different layers. Sloping
dash-doted line gives a slope of the $b(L_x)$ function for ideal tetragonal
lattice.

This plot can be treated as a phase diagram of the system. This diagram is
done in terms of the mean density $\varrho \symbol{126}1/L_xL_y\;$at fixed $%
L_z$ .

Vertical dashed line (at $L_x\approx 11$) denotes a transition between two
different structures. It can be proven that for this case the relation
between two distances (on upper and lower curves) is equal exactly to the $%
\sqrt{3}$. This relation corresponds to ideal 3D hexagonal ordering.

Dotted vertical line corresponds to a transition from a charge order like
Wigner crystals to the phase separation structure which contains charge-rich
droplets with an internal fine structure. It should be noted that the curves
$b(L_x)$ do not reach horizontal asymptotes.

It means that the internal density of the charges in the droplets depends on
the value of averaged density in the whole system. In spite of some analogy
with a picture of liquid droplets this fact is caused by absent of direct
attraction among particles. The particles can not collect together in
compact droplets. The only effective attraction exists here which is caused
by a pairing of those particles from different layers which are close to the
droplet boundaries. As a result, the size of the droplets is always
comparable with a distance between them. From physical point of view the
dependence of the internal density of charge in the droplets from the doping
level is one of the mostly important feature of this new droplet phase.

\section{Acknowledgment}
This work is supported in parts by the Foundation of Fundamental
Research of Ukraine (Agreement F4 72-97, project No 2.4/199) and
INTAS Grant No 96-0410.

\eject

\eject

\eject

\eject

\eject

\eject

\eject

\end{document}